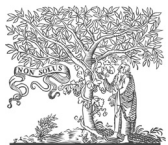
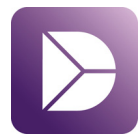

Contents lists available at ScienceDirect

# Data in Brief

journal homepage: www.elsevier.com/locate/dib

Data Article

# A new SWATH ion library for mouse adult hippocampal neural stem cells

Clarissa Braccia [a], Meritxell Pons Espinal [b], Mattia Pini [a], Davide De Pietri Tonelli [b], Andrea Armirotti [a],*

[a] D3 PharmaChemistry, Fondazione Istituto Italiano di Tecnologia, Via Morego 30, 16163 Genova, Italy
[b] Neurobiology of miRNA Lab, Neuroscience and Brain Technologies Department, Fondazione Istituto Italiano di Tecnologia, Via Morego 30, 16163 Genova, Italy

### ABSTRACT

Over the last years, the SWATH data-independent acquisition protocol (Sequential Window acquisition of All THeoretical mass spectra) has become a cornerstone for the worldwide proteomics community (Collins et al., 2017) [1]. In this approach, a high-resolution quadrupole-ToF mass spectrometer acquires thousands of MS/MS data by selecting not just a single precursor at a time, but by allowing a broad $m/z$ range to be fragmented. This acquisition window is then sequentially moved from the lowest to the highest mass selection range. This technique enables the acquisition of thousands of high-resolution MS/MS spectra per minute in a standard LC–MS run. In the subsequent data analysis phase, the corresponding dataset is searched in a "triple quadrupole-like" mode, thus not considering the whole MS/MS scan spectrum, but by searching for several precursor to fragment transitions that identify and quantify the corresponding peptide. This search is made possible with the use of an ion library, previously acquired in a classical data dependent, full-spectrum mode (Fabre et al., 2017; Wu et al., 2017) [2,3]. The SWATH protocol, combining the protein identification power of high-resolution MS/MS spectra with the robustness and accuracy in analyte quantification of triple-quad targeted workflows, has become very popular in proteomics research. The major drawback lies in the ion library itself, which is normally demanding and time-consuming to build. Conversely, through the realignment of chromatographic retention times, an ion library of a given proteome can relatively easily be tailored upon "any" proteomics experiment done on the same proteome. We are thus hereby sharing with the worldwide proteomics community our newly acquired ion library of mouse adult hippocampal neural stem cells. Given the growing effort in neuroscience research involving proteomics experiments (Pons-Espinal et al., 2017; Sarnyai and Guest, 2017; Sethi et al., 2015; Bramini et al., 2016) [4,5–7], we believe that this data might be of great help for the neuroscience community. All the here reported data (RAW files, results and ion library) can be freely downloaded from the SWATHATLAS (Deutsch et al., 2008) [8] website (http://www.peptideatlas.org/PASS/PASS01110)

**Specifications Table**

| | |
|---|---|
| Subject area | *Proteomics, Neuroscience* |
| More specific subject area | *SWATH proteomics* |
| Type of data | *Ion library for SWATH proteomics* |
| How data was acquired | *2D-LC MS/MS in Data-Dependent mode* |
| Data format | *Raw data and a compiled ion library* |
| Experimental factors | *Tryptic digest of mouse adult hippocampal neural stem cells fractionated by 2D LC* |
| Experimental features | *Data dependent MS/MS acquisition followed by database search against Mus Musculus reference proteome.* |
| Data source location | *Fondazione Istituto Italiano di Tecnologia, Via Morego 30, 16163 Genova, Italy* |
| Data accessibility | *All data are freely downloadable from* www.swathatlas.org |

**Value of the data**

- We here report on the building of a new SWATH ion library of mouse adult hippocampal neural stem cells. The whole process from sample preparation to data acquisition to data analysis is carefully described.
- All the data here described are made available to the worldwide community and can be freely downloaded from the SWATH Atlas website (www.swathatlas.org).
- This dataset might represent a ready-made tool for the growing community of SWATH proteomics and it can use to identify and quantify proteins from similar biological contexts

## 1. Data

The data we are hereby sharing is a ready-made ion library to be used for SWATH proteomics experiments by the worldwide proteomics community working on neural stem cells. The library (along with all the corresponding RAW and results files that were used to build it) is openly shared through the SWATHATLAS initiative [8].

## 2. Experimental design, materials and methods

All the chemicals and solvent were purchased from Sigma Aldrich (Milano, Italy), unless otherwise indicated. The NanoAcquity LC system, trapping column and Oasis SPE columns were purchased from Waters (Milford, MA,USA). The 5600+ Triple tof system, ProteinPilot and MarkerView softwares and the PepCalMix reference standard were purchased from SCIEX (Ontario, Canada).

## 3. Sample preparation: aNSC growth, cell lysis and protein digestion

Mouse adult hippocampal neural stem cells (aNSCs) were obtained as previously described [4]. Briefly, adult dentate gyrus (DG) was isolated from 8–10 C57Bl6/J mice at the age of 6–8 weeks. After dissection in Hanks Balanced Salt Solution (HBSS, Gibco) medium, the tissue was enzymatically dissociated with papain (2,5 U/ml), dispase (1 U/ml) and DNaseI (250 U/ml) for 20 min at 37 °C. During incubation, the tissue was repeatedly triturated with a fire polished Pasteur pipette. The cell suspension was centrifuged at 130 g for 5 min and the pellet was re-suspended in a buffered solution (1× HBSS, 30 mM Glucose, 2 mM HEPES pH 7,4, 26 mM NaHCO3) followed by a centrifugation at 130 g for 5 min. aNSCs were isolated using 22% Percoll gradient solution. After further centrifugation for 5 min at 130 g the cell pellet was re-suspended in 2 ml of culture medium containing Neurobasal (Invitrogen), Glutamax (Invitrogen), 1% penicillin and streptomycin (Invitrogen), B27 without retinoic acid (Invitrogen), FGF (20 ng/ml; PeproTech) and EGF (20 ng/ml; PeproTech). The dissociated DG tissue was plated into PDL/Laminin (Sigma/Roche) coated wells and incubated at 37 °C with 5% $CO_2$. The growth medium was exchanged 24 h later to further remove excess debris. Every 2 days half of the growth medium was exchanged with fresh medium to replenish the growth factors. aNSCs were passaged once they reached 80% confluence. aNSCs cell pellets were homogenized using lysis RIPA buffer containing phosphatase and protease inhibitors (Complete mini EDTA-free, Roche) with the following composition: 150 mM NaCl, 1.0% Triton X-100, 0.5% sodium deoxycholate, 0.1% SDS (sodium dodecyl sulphate), 50 mM Tris, pH 8.0 and 1 mM $Na_3VO_4$. Cellular debris were separated at 13,000 rpm for 30 min at 4 °C. The total protein content in the lysate was quantified using BCA assay following the manufacturer instructions (Pierce TM BCA Protein Assay Kit; Thermo Scientific). The volume corresponding to 200 μg of proteins from aNSC was incubated with 10 μl of 100 mM DTT in 50 mM $NH_4HCO_3$, pH 8 for 30 min at room temperature to reduce protein cysteine groups. After reduction, 30 μl of 100 mM IAA in 50 mM $NH_4HCO_3$, pH 8 were added and the sample was incubated for 20 min in the dark. The protein content was then precipitated for 5 h at −20 °C by adding 1 ml of cold (−20 °C) acetone. After centrifugation at 14,000 g for 15 min, the surnatant was discarded and the protein pellet was dissolved in 200 μl of 50 mM $NH_4HCO_3$, pH 8. The sample was then digested overnight at 37 °C by adding 4 μg of trypsin (proteomics grade, from Sigma).

## 4. Desalting and fractionation of peptides

In order to increase the number of identified proteins and their sequence coverage, the resulting tryptic peptides were fractionated offline using the reversed phase high pH/low pH strategy [9]. High pH fractionation was performed using Oasis solid-phase extraction (SPE) columns. After the activation (with 500 μl of ACN) and equilibration (with 500 μl of 3% ACN 0.1% FA) of the column resin, the peptides were loaded on the column. The peptides retained on the column were first desalted with 500 μl of washing solution (3% ACN with 0.1% FA). A total of 8 fractions of peptides were collected by eluting sample with 500 μl of 0.1% TEA solution with increasing acetonitrile concentration (5%, 7.5%, 10%, 12.5%, 15%, 17.5%, 20% and 50%).

## 5. LC–MS/MS analysis

Each fraction was dried in a vacuum centrifuge and then redissolved in 45 µl of 3% ACN added with 0.1% FA. 5 µl of PepCalMix solution, corresponding to 250 pmoles, were added to each fraction as internal standard for retention time recalibration. Five microliters of sample (roughly corresponding to 2.5 µg of peptides) were loaded on a NanoAcquity chromatographic system and connected to a TripleToF 5600+ mass spectrometer equipped with a NanoSpray III ion source. The peptides were first trapped on a 180 µm × 20 mm Acquity C18 column for 4.0 min at 4.0 µl/min flow rate. Eluent composition was kept at 1% ACN + 0.1% FA. Peptides were then moved on a PicoFrit C18 column (75 µm × 25 cm, from NewObjective Inc., Woburn, MA, USA) and eluted at 300 nL/min with a 2 h gradient of acetonitrile in water (3–45%, both eluents were added with % FA). ACN content was then ramped at 90% in 5 min and kept for additional 5 min. The system was then reconditioned to 3% for 18 min. Eluted peptides were analyzed in positive ion mode. Ion spray voltage was set to 2400 V, spray gas 1 and curtain gas values were set to 4 and 30 respectively. Source temperature was kept at 75 °C and declustering potential was set to 80 V. Up to 40 precursor ions were selected for MS/MS experiment using the following criteria: charge state +2 to +5 with intensity above 150 counts. Scan range was set to 400–1200 $m/z$ for the MS survey scan and to 100–1600 $m/z$ for each MS/MS scan. Collision energy values recommended by SCIEX were used for peptide fragmentation. The total ion chromatograms for the eight fractions is represented in Fig. 1.

## 6. Protein identification and creation of the ion library

A total of 313362 MS/MS spectra were collected in our experiment (an average of 43 MS/MS spectra per second, considering the whole gradient time). Raw data files from all the fractions were analyzed using ProteinPilot and MASCOT (Matrix science Ltd, London, UK) software. In both cases, MS/MS spectra were searched against the reference *Mus Musculus* proteome downloaded September the 27, 2017 from UNIPROT database (http://www.uniprot.org/proteomes/UP000000589) and consisting of 52015 proteins. Carbamidomethylation was set as fixed modification. The purpose of this effort was to ensure the highest quality to our ion library, by only retaining proteins at maximum 1% false discovery rate (FDR). FDR analysis is normally performed by searching MS/MS data against a given target database and against a corresponding decoy database [10], where the target sequence is randomized (same aminoacid composition, but randomized sequence) or semi-randomized (same aminoacid composition, correct first and last residues, randomized remaining sequence). The ratio between the number of protein IDs found against the decoy database and the number of protein IDs found against the target database returns the FDR estimate for that dataset (Global FDR) [11]. ProteinPilot, working with the Paragon algorithm [12], uses a *confidence score* to assign identification likeliness at peptide spectrum match level (PSM FDR) and it provides a calculation of the corresponding FDR value at protein level by using the method published by Tang [13]. A preliminary search of our data with ProteinPilot, using a minimal set of PTMs (Rapid search, consisting of methionine oxidation, asparagine and glutamine deamidation and conversion of glutamine and glutamate to pyroglutammic acid) and setting a 10% confidence score threshold (default search parameters recommended by SCIEX for optimal FDR calculation) returned **4394** IDs having a calculated maximum FDR of 1%. The algorithm also estimated that a protein FDR of 1% would correspond to a confidence score of 79.6%. Based on this estimate, a dataset obtained with 80% confidence threshold should already consist of proteins having 1% maximum FDR. We then refined our searches using ProteinPilot at increasing confidence scores (20–99.9%) and we plotted the corresponding number of identified proteins. These numbers of decoy proteins were found by ProteinPilot at each different confidence score: 20%: 879, 60%: 165, 80%: 90, 90%: 24, 99%: 1, 99.9%: 0. As a comparative test, we then searched our data using MASCOT software with only one PTMs (methionine oxidation), in order to reduce the overall chance of false hits. MASCOT allows an explicit control of FDR at protein level and it returned **3779** protein hits at 1% FDR protein level. The results of the search are shown in Fig. 2.

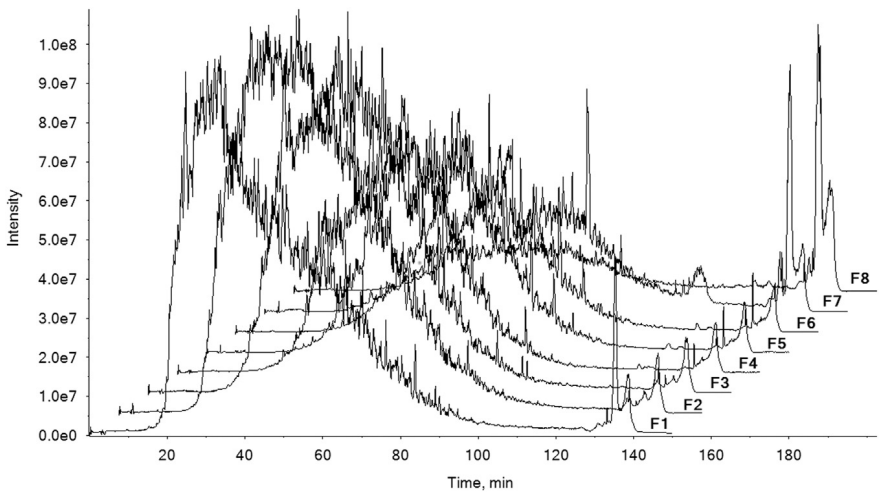

**Fig. 1.** Overlapped LC-MS traces of the eight fractions from high-pH fractionation.

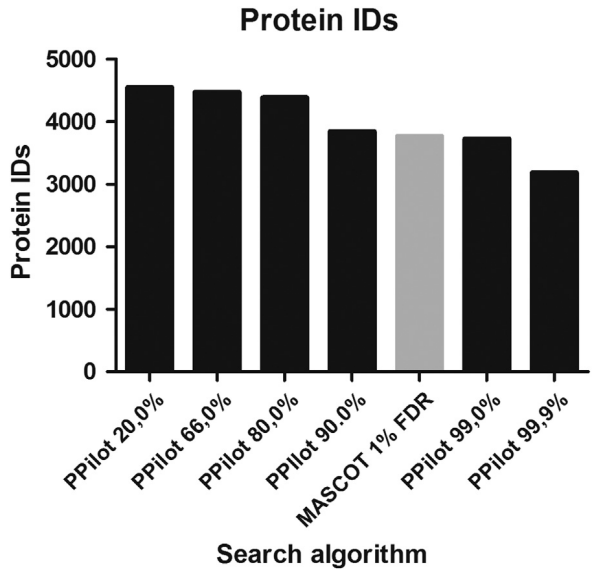

**Fig. 2.** Number of protein IDs at different ProteinPilot confidence scores (black) compared with MASCOT protein IDs at 1% FDR (grey).

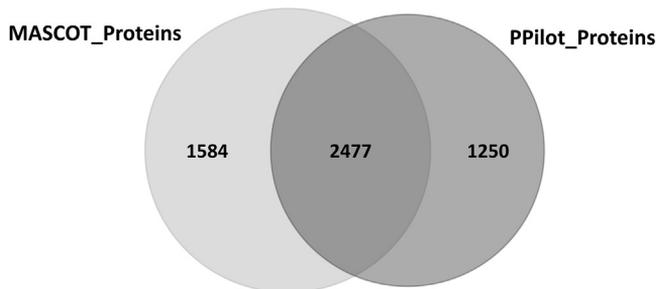

**Fig. 3.** Venn diagram showing the overlapping of the two dataset at individual protein level.

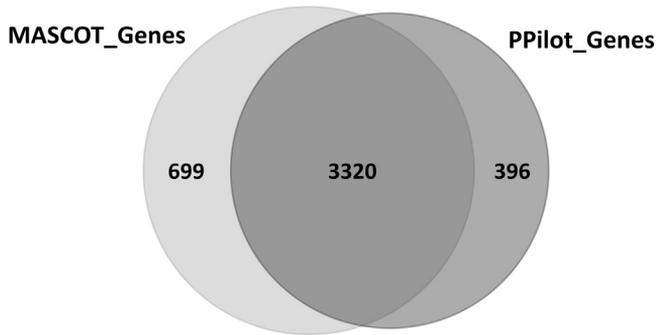

**Fig. 4.** Venn diagram showing the overlapping of the two dataset at gene level.

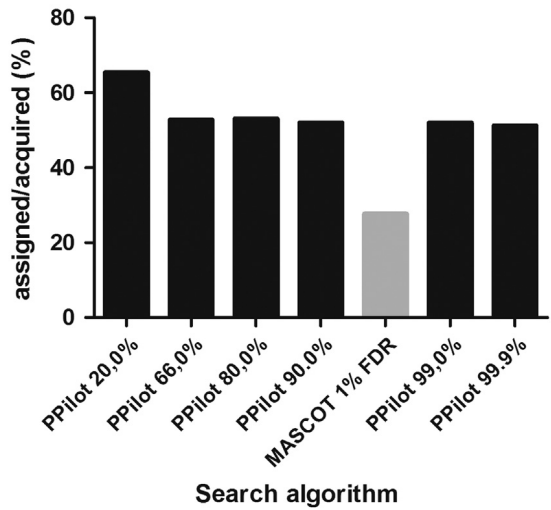

**Fig. 5.** Percentage of MS/MS spectra positively assigned to protein hits by MASCOT (grey) and ProteinPilot at different confidence scores (black).

A ProteinPilot search done with a confidence score of 99% returns slightly *less* protein IDs than a MASCOT search with a controlled protein 1% FDR (3727). We then checked the degree of overlapping of the two sets of results, by plotting the corresponding Venn diagrams at the level of both individual proteins (Fig. 3) and corresponding genes (Fig. 4)

As demonstrated by the plots, the overlapping degree of the two dataset is not particularly high at individual proteins level, with roughly 65% ProteinPilot IDs shared with MASCOT. The degree of overlapping is much higher at the level of the corresponding genes, with roughly 90% of the genes represented by ProteinPilot protein IDs shared with MASCOT. The conversion of the protein IDs into their corresponding genes was done using the ID mapping tool on the Uniprot website (http://www.uniprot.org/uploadlists/). We also investigated the role of the two algorithms (Paragon and MASCOT) in the total number of MS/MS spectra positively assigned to protein hits (Fig. 5).

The number of MS/MS spectra assigned by MASCOT to protein hits, compared to the total number of acquired ones (313362) is markedly lower (28%) than those assigned by ProteinPilot (51–53%). Besides the differences between the two search algorithms, this difference is also due to lower number of PTMs used for MASCOT (1) compared to ProteinPilot (5). The latter software does not allow the selection of individual PTMs in the search other than those indicated above for the "Rapid" search effort. This discrepancy in PTMs might also partially explain the differences with MASCOT in the overlapping of individual proteins compared to the corresponding genes (Figs. 3 and 4).

Table 1
Features incorporated in the aNSC ion library.

| | |
|---|---|
| MRM Assays | 594922 |
| Peptides | 25394 |
| Proteins (Max 1% FDR) | 3673 |

## 7. Creation of the ion library and performance test

We built the SWATH ion library applying the most stringent quality criteria we could. The.group file resulting from Protein Pilot search at 99% confidence was loaded on the SWATH microapp tool embedded in PeakView algorithm. No peptides shared by more than one protein were imported. We then removed from the ion library all the peptides carrying PTMs other than CAM, thus allowing only this fixed modification. The reference data (transitions and expected RT values) for the PepCalMix were manually added the ion library using a text editor. Table 1 reports a summary of the features of the final ion library

For each entry, the library reports the Q1 and Q3 $m/z$ values (precursor and fragment respectively), the observed retention time (in minutes), the relative intensity of the transition, the associated protein name and peptide sequence, the precursor charge state and the fragment type, plus several other flags. The above described ion library has been uploaded in the SWATATLAS website and can be freely downloaded from http://www.peptideatlas.org/PASS/PASS01110. Any TripleToF user worldwide can immediately use the ion libraries we are hereby sharing. Commercially available standard peptides for retention time recalibration other than the PepCalMix sold by SCIEX can in principle be used. The ion library must then be updated with the corresponding MRM and RT data. This can easily be done by using a text editor.

## Acknowledgements

The authors wish to thank Dr. Zuzana Demianova (SCIEX) for helpful discussions and Dr. Caterina Gasperini for maintenance of aNSCs culture. This work has received funding from the European Union's Horizon 2020 research and innovation programme under Grant agreement no. 696656 – Graphene Core 1.

## Transparency document. Supplementary material

Transparency document associated with this article can be found in the online version at http://dx.doi.org/10.1016/j.dib.2018.02.062.